\documentclass[final,5p,times,twocolumn]{elsarticle}

\usepackage{amssymb}
\usepackage{amsthm}
\usepackage{amsmath}
\usepackage[ruled,vlined]{algorithm2e}
\usepackage{hyperref}

\newcommand{\vect}[1]{\boldsymbol{#1}}

\newcommand{\cT}{\vect{\mathcal{Q}}} 

\usepackage[dvipsnames]{xcolor}

\journal{Future Generation Computer Systems}

\begin{document}

\begin{frontmatter}


\title{Advancing Anomaly Detection in Computational Workflows with Active Learning}

\author[anl]{Krishnan Raghavan}
\ead{kraghavan@anl.gov}

\author[isi]{George Papadimitriou\corref{cor1}}
\ead{georgpap@isi.edu}

\author[anl]{Hongwei Jin}
\ead{jinh@anl.gov}

\author[renci]{Anirban Mandal}
\ead{anirban@renci.org}

\author[ornl]{Mariam Kiran}
\ead{kiranm@ornl.gov}

\author[ornl]{Prasanna Balaprakash}
\ead{pbalapra@ornl.gov}

\author[isi]{Ewa Deelman}
\ead{deelman@isi.edu}

\cortext[cor1]{Corresponding address: USC Information Sciences Institute, 4676 Admiralty Way Suite 1001, Marina del Rey, CA, USA, 90292}

\address[isi]{University of Southern California, Information Sciences Institute, Marina del Rey, CA, USA}

\address[renci]{RENCI, University of North Carolina Chapel Hill, NC, USA}

\address[anl]{Data Science and Learning Division, Argonne National Laboratory, IL, USA}

\address[ornl]{Oak Ridge National Laboratory, TN, USA}

\begin{abstract}
A computational workflow, also known as workflow, consists of tasks that are executed in a certain order to attain a specific computational campaign. Computational workflows are commonly employed in science domains, such as physics, chemistry, genomics, to complete large-scale experiments in distributed and heterogeneous computing environments. However, running computations at such a large scale makes the workflow applications prone to failures and performance degradation, which can slowdown, stall, and ultimately lead to workflow failure. Learning how these workflows behave under normal and anomalous conditions can help us identify the causes of degraded performance and subsequently trigger appropriate actions to resolve them. However, learning in such circumstances is a challenging task because of the large volume of high-quality historical data needed to train accurate and reliable models. Generating such datasets not only takes a lot of time and effort but it also requires a lot of resources to be devoted to data generation for training purposes. Active learning is a promising approach to this problem. It is an approach where the data is generated as required by the machine learning model and thus it can potentially reduce the training data needed to derive accurate models. In this work, we present an active learning approach that is supported by an experimental framework, Poseidon-X, that utilizes a modern workflow management system and two cloud testbeds. We evaluate our approach using three computational workflows. For one workflow we run an end-to-end live active learning experiment, for the other two we evaluate our active learning algorithms using pre-captured data traces provided by the Flow-Bench benchmark.
Our findings indicate that active learning not only saves resources, but it also improves the accuracy of the detection of anomalies.
\end{abstract}




\end{frontmatter}

\section{Introduction}

Computational workflows are abstractions widely used to represent complex computational processes~\cite{da2021workflows}. Workflows describe the computational tasks to be executed, the relative order in which they have to be executed, and how data movements need to be performed. Workflows have played a key role in advancing science by allowing scientists to
orchestrate computations at massive scales, from cancer research~\cite{wozniak2018candle} 
to molecular dynamics~\cite{sivaraman2020machine}. 
However, executing workflows at large scale in distributed and heterogeneous computing environments makes them more likely to experience failures and performance degradation. Failing to identify and determine the cause of such anomalies in time, can be detrimental to the overall efficiency and performance of computational workflows.

Machine learning approaches have emerged as a viable solution to the anomaly detection problem.
Recent ML driven anomaly detection approaches include deep learning methods~\cite{krawczuk-pearc-2021} and graph neural network approaches~\cite{jin2023graph} to identify anomalies in computational workflows. These approaches, even though they produce models that perform well within the scope of their studies, are data-hungry, and require high volumes of high-quality labeled training data that cover a broad spectrum of scenarios. However, the lack of publicly available datasets that (1) include labeled anomaly data (2) correlate application and workflow events with infrastructure performance traces and (3) provide DAG structure information, has led prior works~\cite{krawczuk-pearc-2021}\cite{jin2023graph} to generate new datasets using custom approaches to be able to conduct their studies. The generation of such datasets necessitates the design of new experiments, continuous deployment and constant evaluation of the data quality. This is not only time-consuming and wastes a lot of resources, but it is also labor intensive and susceptible to data quality issues.

In this work, we make strides towards addressing the data generation problem, within the context of workflow performance traces, and the need for large volume of training data to produce accurate anomaly detection ML models. First we develop an experimental framework, Poseidon-X, that leverages a workflow management system and two NSF-funded cloud testbeds to provide reproducible and on-demand data collection of computational workflow executions. To develop this framework we have chosen Pegasus~\cite{deelman-cise-2019} as the workflow management system of choice, while compute and network resources are being orchestrated on the Chameleon Cloud~\cite{chameleon} and the FABRIC testbed~\cite{fabric-baldin-2019}.
This approach enables Poseidon-X to support data collection from any workflow, after being ported into a Pegasus workflow. Additionally, by using a custom isolated deployment across the two cloud testbeds, Poseidon-X supports injection of different types of anomalies at variable magnitudes, which are tracked automatically and are correlated with the workflow executions to generate quality labeled datasets.
Then we develop a novel active learning approach for computational workflows, and we demonstrate that an unsupervised learning model can be trained incrementally by guiding the generation of experimental data on the fly. To address the need for unsupervised learning, we utilize a self-supervised learning model (SSL), which was presented in~\cite{jin2023ssl}. To train this SSL model, instead of traditional random data sampling strategies, we develop a feedback loop. Over successive iterations, the model is trained using data generated through on-demand workflow executions (live experimentation), that take place on the Poseidon-X framework.
In our approach, we drive the workflow experiments by measuring the confidence of the model on the training data and we use leads from the low-confidence regions to trigger new targeted experiments. The new experiments aim to improve our knowledge of these data segments. We then retrain the model with the newly acquired data to maximize the model's confidence.

\noindent The contributions of this paper are twofold:

\begin{enumerate}
    \item We develop an experimental framework that can be used to model any computation workflow and automatically collect and label performance data from their executions within a reproducible and on-demand environment.
    \item We develop the mathematical framework required for active learning and sampling efficiently in the unsupervised learning setting. We present a methodology that uses a feedback loop that provides the model with newly generated data, through the aforementioned experimental framework, driven directly by the model's confidence levels.
\end{enumerate}

\noindent The rest of this paper is organized as follows: in Section~\ref{sec:related} we present related work, in Section~\ref{sec:methodology} we present our experimental framework, Poseidon-X, and active learning methodology, in Section~\ref{sec:experiments} we present experimental results and in Section~\ref{sec:conclusion} we conclude with our findings and future work.

\section{Related Work}
\label{sec:related}
Current literature is abundant with anomaly detection methods. However, a majority of these methods have been developed for datasets in the form of images and tabular data, see \cite{samariya2023comprehensive} for a survey. For instance, common anomaly detection methods include clustering~\cite{sohn2023anomaly}, principal component analysis (PCA)~\cite{ma2023mppca}, and auto-regressive integrated moving average models~\cite{goldstein2023special}. Even though these methods are efficient for images and tabular data, they cannot be applied directly in the case of graphs, because of the need to capture the structural information in graphs.

More recently, numerous machine learning (ML) methods have been developed for anomaly detection in graphs. Recent application~\cite{ouyang2020unified, feng2023unsupervised, kim2022graph} indicates that graph neural networks (GNNs)  can be very effective for anomaly detection because they explicitly consider structural information presented by graphs. Since, workflow data is represented by graphs, GNN-based anomaly detection methods are rather appropriate and the first known application of GNN in workflow anomaly detection was demonstrated in~\cite{jin2022workflow, jin2023graph}. Albeit, in the supervised learning scenario, impressive results were presented. 

The key requirement of supervised machine learning is the need for labeled data from workflow. When considering large-scale heterogeneous infrastructure, labeled data is extremely tedious to generate.  To obviate this necessity, Jin et al.~\cite{jin2023ssl} introduced a self-supervised learning approach where the workflow data is labeled during the learning process. However, even with the reduced requirement of labeling, the necessity to generate large quantities of data prior and the large-scale experiments needed for these purposes are still a considerable drain on the resources. Therefore, it is rather desirable to generate the required amount of data using feedback from the ML model, in order to make the overall sampling process more efficient, yet still accurate.

Active learning is a paradigm aimed at improving the sample efficiency of machine learning models by efficiently and appropriately sampling data guided by the uncertainty and confidence of the model. Recent applications of active learning for autonomous experimentation in scientific domains and experimental design have indicated enormous potential~\cite{ren2023autonomous}.
Presently, active learning methods only cater to experimental design~\cite{ren2023autonomous} on datasets involving 
numerical data~\cite{sauer2023active} or images~\cite{giancola2023towards}.
To our knowledge, no implementation of active learning in the domains of datasets involving graphs exists. Moreover, active learning has never been demonstrated either in the workflow management domain or in the unsupervised anomaly detection domain. 


The paradigm of active learning requires the infrastructure for on-the-fly generation of data. Many efforts have focused on providing back-ends to generate new data for computational workflows. These solutions include emulation and simulation approaches, as well as live experimentation on purpose-specific testbeds. Simgrid~\cite{casanova:hal-01017319} is a simulation framework for distributed computing capable of simulating complex network topologies, HPC clusters, cloud deployments, and more. Wrench~\cite{wrench} is a cyberinfrastructure simulation framework tailored for distributed computing applications, systems, and platforms. It relies on SimGrid and has been used successfully for research, development, and education. Both of them can provide fast simulation times, but their accuracy depends on the application and careful calibration. Testbeds enable real experimentation in a controlled environment. For example,
Aerial Experimentation and Research Platform for Advanced Wireless (AERPAW)~\cite{aerpaw-wintech-2020}\cite{aerpaw-dronet-2022} provides an emulation module and infrastructure to conduct live experiments in workflows that control drone operations. 

In summary, there is a need to develop experimental infrastructure and efficient interfacing along with an active learning framework to generate data on the fly and efficiently detect anomalies in workflow. To this end, we build an experimental infrastructure using two NSF-funded testbeds, and we extend the SSL anomaly detection model presented in~\cite{jin2023ssl} to the active learning paradigm. Our approach demonstrates effective learning while performing anomaly detection on workflows. To the best of our knowledge, no active learning approach using feedback from real infrastructure has been demonstrated with graphs or in the context of workflow management systems.

\section{Overall Methodology}
\label{sec:methodology}
\begin{figure*}[th]
\centering
\includegraphics[width=\linewidth]{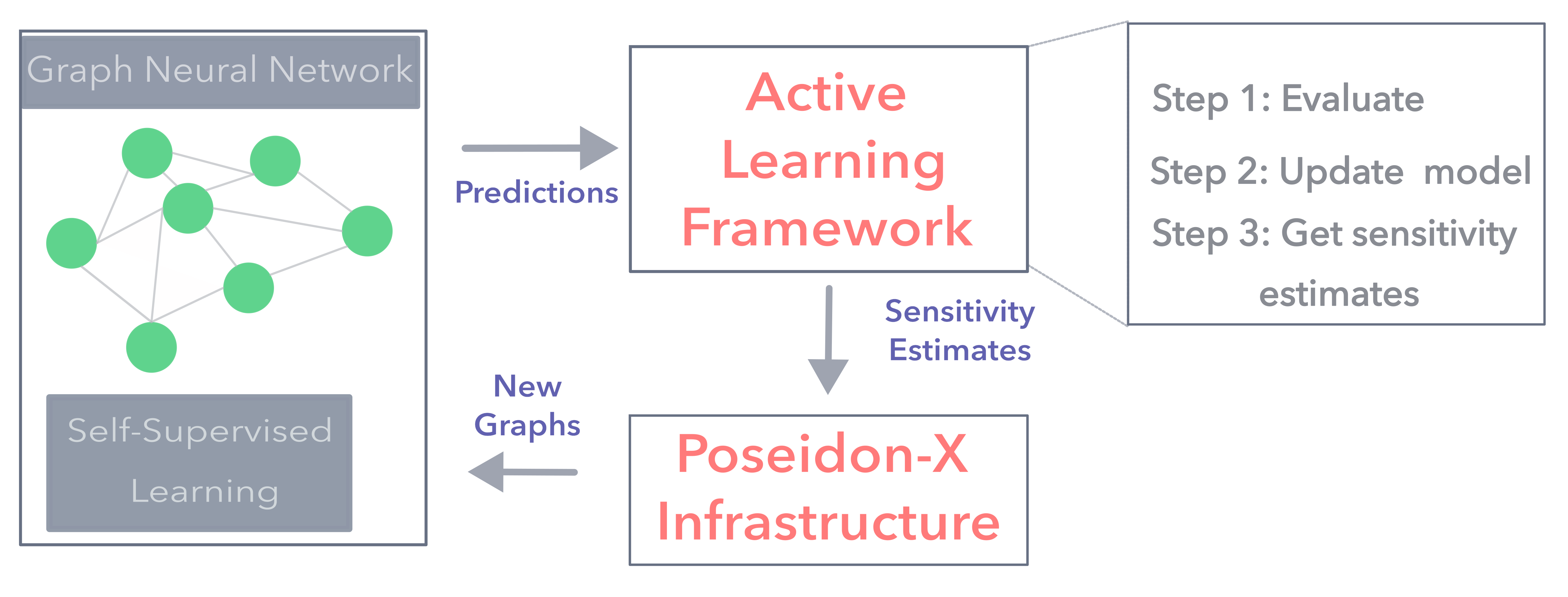}
\caption{Overview of the end-to-end active learning framework. This framework has three modules, the neural network (graph neural network or self-supervised learning), the active learning module and the Poseidon-X infrastructure. The active learning module trains the network and passes sensitivity scores to the Poseidon-X infrastructure which then generates the requisite data. }
\label{fig:overview}
\end{figure*}

There are three components of our framework as illustrated in Figure \ref{fig:overview}, the experimental infrastructure-- Poseidon-X module that generates the data and executes experiments on the fly, the active learning module, and the neural network, either the graph neural network or the  self-supervised learning-driven neural network.
\subsection{Experimental Infrastructure}
\label{sec:experimental-infra}
To support active learning we need a robust experimental harness that can drive workflow execution experimentation on a real infrastructure and can automatically collect and process performance statistics. With this goal in mind, we built Poseidon-X using a highly programmable cloud infrastructure, a well-established workflow management system, and a newly developed experiment controller.

\subsubsection{Cloud Infrastructure}
Poseidon-X is backed by two National Science Foundation (NSF) funded cloud testbeds.

\noindent\textbf{FABRIC}~\cite{fabric-baldin-2019}, is a nationwide instrument that enables large-scale experimentation within an isolated, yet realistic environment. It provides compute and storage resources on multiple sites that are interconnected by high-speed, dedicated optical links. FABRIC also offers ``everywhere programmability'', as all of its resources can be adapted and customized for each experiment's needs.

\noindent\textbf{Chameleon Cloud}~\cite{chameleon}, is a large, deeply programmable testbed, designed for systems and networking experiments. It leverages OpenStack~\cite{openstack} to deploy isolated slices of cloud resources for user experiments. However, where FABRIC scales in geographic distribution, Chameleon scales by providing large amounts of compute, storage, and networking resources spread across six sites. Users can provision bare metal compute nodes with custom system configurations connected to user-controlled OpenFlow~\cite{openflow} switches. In addition, Chameleon networks can be connected to external partners and other facilities, including FABRIC slices, using the concept of facility ports. 

Finally, both testbeds offer Python APIs to control their resources. FABRIC offers the \texttt{FABlib} API and Chameleon offers the \texttt{python-chi} API, which we use to design reproducible active learning experiments.

\subsubsection{Pegasus Workflow Management System}
Pegasus~\cite{deelman-cise-2019} is a popular workflow management system that enables users to design workflows at a high level of abstraction that describes the workflows in a way that is independent of the resources available to execute them and also independent of the location of data and executables.
Workflows are described as directed acyclic graphs (DAGs), where nodes represent individual compute tasks and the edges represent data and control dependencies between tasks.
Pegasus transforms these abstract workflows into executable workflows that can be deployed onto distributed and high-performance computing resources, shared computing resources, local clusters, and clouds. 
Through its Panorama extensions~\cite{pegasus-panorama},~\cite{papadimitriou-fgcs-2021}, Pegasus also offers end-to-end online workflow execution monitoring and provides execution traces of the computational tasks, statistics for individual transfers and other infrastructure related metrics.  This information gets stored in an Elasticsearch instance~\cite{elastic} for ease of access and analysis.

Within Poseidon-X, we use Pegasus to execute scientific workflows on the cyberinfrastructure deployed in FABRIC and Chameleon and collect statistics and logs to train our models.

\begin{figure*}[t]
\centering
\includegraphics[width=\linewidth]{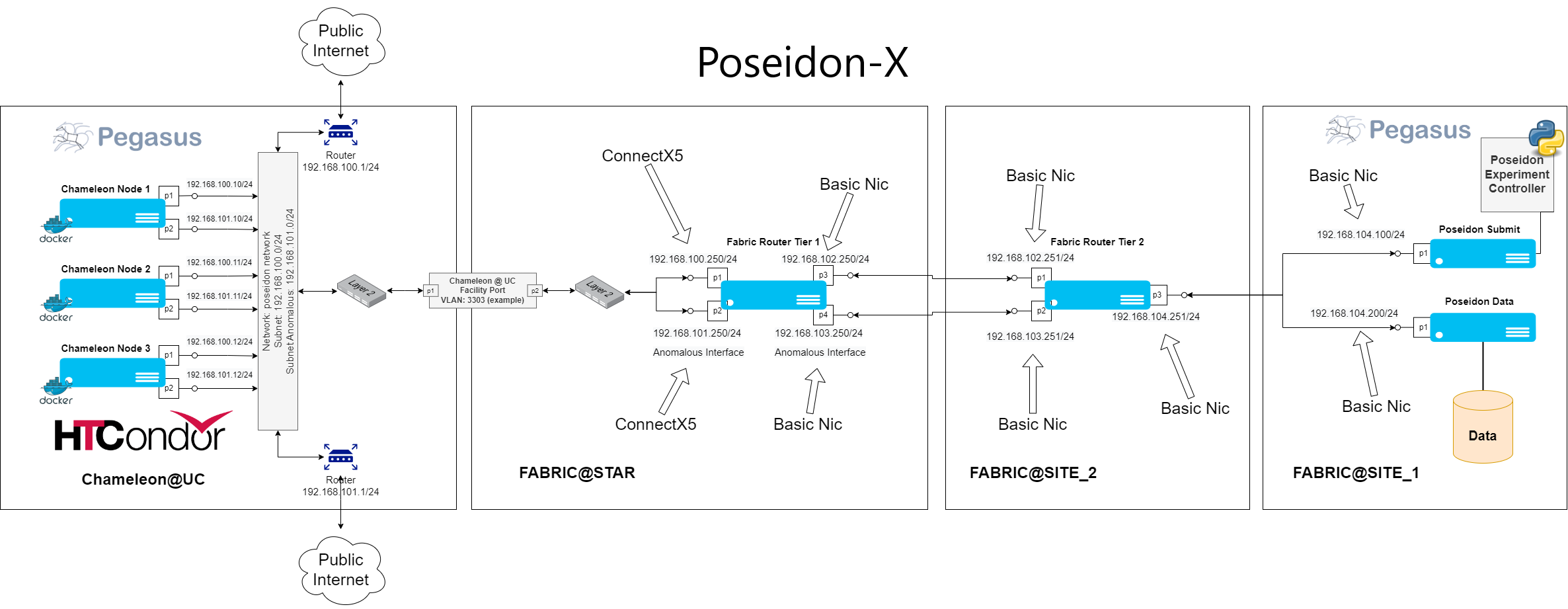}
\caption{Poseidon-X experimental infrastructure that operates as the back-end of our active learning framework. Poseidon-X spans across the Chameleon Cloud and the FABRIC testbed. Chameleon hosts the workers while FABRIC host the networking infrastructure, the workflow submission node and the data storage node. To use the resources efficiently Poseidon-X uses Docker containers on baremetal nodes and injects CPU and HDD interference on selected containers using cgroups. To model network anomalies, flows are being directed faulty paths in FABRIC. An experimental controller on the workflow submission node orchestrates the anomaly injection, workflow execution triggering and data labelling.}
\label{fig:poseidon-deployment}
\end{figure*}

\subsubsection{Experiment Controller}
Since active learning requires many rounds of experiments to be conducted, with different settings, an experiment controller is required to streamline the process. We have designed and developed a Python-based controller that receives as input (1) a JSON description of the available infrastructure that has been provisioned from the testbeds, (2) a JSON description with the list of experiments that need to be conducted, and (3) a JSON description of the available workflows, for which the controller can orchestrate experiments. For each experiment, the controller deploys an HTCondor~\cite{htcondor} pool, by starting HTCondor workers within Docker~\cite{docker} containers in Chameleon. After the pool stabilizes, the experiment controller starts and monitors the workflow execution. During execution, Pegasus collects logs and measurements into a performance database and Elasticsearch. After the workflow execution successfully is completed, the experiment controller parses the logs and creates a single CSV file with aggregated statistics.

\subsubsection{Anomaly Injection}
Poseidon-X supports the introduction of slowdowns in computing resources and it also supports network anomaly injections. To introduce slowdowns in computing resources we are leveraging \texttt{cgroup} limits. For example, to introduce CPU slowdown we restrict the workflow binaries to use only a certain number of cores available to the host, and to introduce filesystem slowdowns we limit the average read and write speed. To introduce network anomalies, we use the \texttt{TC} tool on the router that connects the Chameleon cloud with FABRIC, and we apply rules on all the traffic that originates from or is destined to the anomalous Chameleon subnet. These rules can introduce packet loss, packet duplication and other network anomalies.

\subsection{Active Learning  Module}\label{ref:alm}

In this work, our goal is to automate and guide the data-generating process such that any neural network model can request data on the fly. In this section, we develop the mathematical framework that allows us to perform such learning. We will first assume a model $f$ parameterized by weight parameters$\vect{w}.$ For the sake of this paper, this will be a self-supervised learning model or a supervised learning graph neural network. However, any architecture/model can be used within our framework. We will assume that this model is trained iteratively for $K$ iterations, indexed by $k \in {1,2,3,\ldots,K}.$ At each iteration, SSL is provided with a batch of data that is generated by a live experiment. This live experiment creates data to delineate the "uncertain region of the sample space," referring to the area where SSL exhibits the least effective predictive performance. The SSL model determines these uncertain regions by relying on a score $s_{k}.$ We will consider each batch of data to be a task $\cT_{k}.$ This batch of data is represented by a graph that is denoted by a tuple of adjacency matrix $\phi$ and the feature vector $\vect{X}$ such that $\cT_{k}= (\phi, \vect{X})_{k}.$ Given this task $\cT_{k},$ the problem of active learning is to determine the next task $\cT_{k+1}$ such that the new task contributes positively towards the improvement of anomaly detection. There are three steps to the problem; (a) given a task $\cT_{k}$ update the model $f$; (b) generate the score $s_{k}$; and (c) interpret the score to generate the next task $\cT_{k+1}.$ These three steps will be executed repeatedly until a pre-ascertained condition on the score is satisfied. 

Given the goal to minimize sample and resource usage in the learning process, each task will only partially represent the underlying data distribution. Consequently, it is crucial that the model remains unbiased towards any single task and retains knowledge gained from previous iterations. Simultaneously, the model must also embrace new experiences as new data is introduced. This aspect of the active learning strategy introduces a trade-off between retaining old information and assimilating new experiences. Within the realm of continual learning, this manifests as the generalization-forgetting trade-off, a concept formalized in~\cite{raghavan2021formalizing} within a dynamic programming framework that considers the cumulative impact of each task during learning. In our study, we leverage this learning strategy by explicitly modeling the cumulative effect of each new data batch on the learning process.

\subsubsection{Iterative Active Learning}
Without loss of generality, we adopt a loss function corresponding to the model \( f \) at each iteration \( k \). This loss function is expected to be general and should adhere to the common assumptions of twice differentiability prevalent in machine learning literature. With the objective of anomaly detection in mind, we aim to minimize the loss mathematically across all tasks following the continual learning framework introduced in~\cite{raghavan2021formalizing}, expressed as:
\begin{align} \label{eq:opti_problem}
V_k^{(*)} = \underset{\vect{w}}{min} \sum_{\tau=k}^{K} \underset{p(\cT_{\tau})}{\mathbb{E}}  \left[ J_{\tau}( \vect{w} ) \right],
\end{align}
Here, \( J_{\tau}(\mathbf{\theta}) \) represents the cost associated with learning on each task. It's worth noting the resemblance between the cost function defined in the standard machine learning setting and the expression \( \sum_{\tau=k}^{K} \beta_{\tau} \mathbb{E}_{p(\mathcal{T})}  \left[ J_{\tau}( \mathbf{w} ) \right] \). In the standard setting, the optimization problem is tackled for each data batch, i.e., for each \( \tau \), and the cumulative effect is observed through stochastic gradient descent updates on the weights \( \mathbf{w} \). Due to the linearity of updates, the gradient values at each \( k \) are aggregated, as demonstrated by \( \mathbf{w}_{K} = \mathbf{w}_{0} + \sum_{\tau = 0}^{K} \nabla_{\mathbf{w}_{\tau} }  \mathbb{E}_{p(\mathcal{T})}  \left[ J_{\tau}( \mathbf{w} ) \right] \). In contrast, the optimization problem in Equation \ref{eq:opti_problem} explicitly accounts for the cumulative effect. This modeling approach ensures that the cumulative effects of the gradient are taken into consideration when generating the next data batch, thereby minimizing bias against any individual data sample.

As observed, for each batch of data, the optimization problem in Equation \ref{eq:opti_problem} is heavily dependent on the choice of $\underset{p(\cT_{\tau})}{\mathbb{E}}  \left[ J_{\tau}( \vect{w} ) \right]$, which in turn relies on the choice of $\cT_{\tau}.$ Thus, there are two steps of this problem, first determine the sample $\cT_{\tau}$ through the experiment and then solve Equation \ref{eq:opti_problem} to get your weights.
This provides the active learning optimization problem as 
\begin{align} \label{eq:active_learning}
V_k^{(*)} &= \underset{\vect{w}}{min} \sum_{\tau=k}^{K} \beta_{\tau} \mathbb{E}_{\cT_{k}}  \left[ J_{\tau}( \vect{w} ) \right],\\
&\text{such that} \nonumber \\
& \cT_{k} = g(s_{k-1})
\end{align}

However, to solve the problem in Equation \ref{eq:active_learning}, we would need the data samples from $k+1$ to $K$ which is unavailable at the iteration $k.$ We therefore perform algebraic simplifications to write an optimization problem at each step. This is obtained by applying dynamic programming principles with the Taylor series to create a one-step recursive equation, that can be solved recursively to obtain a solution to the active learning problem. Towards this end, we obtain~(see~\cite{raghavan2021formalizing} for a complete derivation) 
\begin{align} \label{eq:first_diff}
     V^{(*)}_{k+1} - V^{(*)}_{k}  &= \\
	- \underset{\vect{w}_{k} \in \Omega_{\theta}}{min} \big[ \beta_k J_{k}(\vect{w}_{k}) 
     +  \langle \nabla_{\vect{w}_{k}} V_{k}^{(*)} , \Delta \vect{w}_{k} \rangle  &+ \langle \nabla_{\cT_{k}} V_{k}^{(*)}, \Delta \cT_{k} \rangle  \big]. \nonumber\\ 
     &\text{such that} \nonumber \\
 \cT_{k} &= g(s_{k-1}) \nonumber
\end{align}
where $\langle a,b \rangle$ describes the inner product between $a$ and $b.$
In Equation \ref{eq:first_diff}, $V^{(*)}_{k+1} - V^{(*)}_{k} $ is the change in the best solution to the active learning problem and represents, ``how the new batch of data, the subsequent parameter update affect the solution to the active learning problem. In particular, this effect is quantified by three terms: the cost contribution of all the information available presently~$J_{k}( {\vect{\theta}}_{k});$ the correlation between the new task and the solution to the active learning problem; the correlation between the optimal cost and the parameters.

The active learning problem is solved when $V^{(*)}_{k+1} - V^{(*)}_{k}$ is small, which is only possible when all three terms on the right-hand side are minimized. The right-hand side is a sum of three terms where the first term $J$ is always positive and the other two terms are not always positive. In essence, we seek to obtain a zero sum~(minimum sum) between the three terms on the right-hand side by selecting sampling $\cT_{k}.$ 

We generate samples that represent all the uncertain regions of the probability space, i.e., the regions that the model is unsure about and does not predict well. We argue that the uncertain regions in the probability space are represented by the impact they have on the solution. In this scenario, the larger the impact presented by the sample to the solution of the active learning problem, the more uncertain we are about this region, and then, this region needs to be explored and learned. 

The insight illuminates that, ``we would like to generate samples that maximize the change in solution to the active learning problem." A robust method of sampling  $\cT_{k}$ that also pertains to the exploration in the probability space $p(\cT_{k})$ is to maximize the impact of $\cT_{k}$ on the three terms in the right and then minimize the impact using the weights of the network, this will ensure that the learning is robust under worst case conditions introduced by the data and the samples are generated to represent the most uncertain regions of the probability space. Mathematically, these regions can be achieved by observing the steepest ascent directions of the term $ \langle \nabla_{\cT_{k}} V_{k}^{(*)}, \Delta \cT_{k} \rangle $ that summarizes the impact of the task on the solution to the active learning problem.

Formally, we may rewrite the optimization problem as a two-step expectation minimization problem such that 
\begin{align} \label{eq:em_act}
 \underset{\vect{w}}{min} & \quad H(\vect{w}, \cT_{k}) \nonumber \\  \text{such that} & \quad \cT_{k} =g(s_{k-1})
\end{align}
where $s_k = \frac{d H(\vect{w}, \cT_{k})  }{d \cT_{k}}$ provides the steepest ascent directions with $H(\vect{w}, \cT_{k})  $ approximated through finite difference approximation such as
\begin{equation}
	\begin{aligned}
		H(\vect{w}, \cT_{k})  &\approx \mathbb{E}_{\cT_{k}} [  \beta_k J_{k}(\vect{w}) + \\ 
        &+( J_{k}(\vect{w} + \Delta \vect{w}) - J_{k}(\vect{w}) ) + \nabla_{\cT_{k}} J(w)],
	\end{aligned}
	\label{eq:approximation}
\end{equation}
where $\zeta$ refers to the number of updates used for the finite difference approximation~(see \cite{raghavan2021formalizing} for a complete derivation of this approximation).
While we solve the minimization step with a standard gradient learning mechanism, the expectation step is solved by generating data from the experiment.

\subsubsection{The data generation mechanism}
To generate data we are using the experimental approach described in Section~\ref{sec:experimental-infra}. Figure~\ref{fig:poseidon-deployment} presents the overall deployment, which spans across two testbeds, and the \texttt{submit node} on $Fabric@site\_1$, where Poseidon-X is deployed trains the machine learning models and controls the execution of the workflow experiments. This is done to enable communication between the active learning module and the experimental controller using files. We have created a translation layer that receives the output from the machine learning model and translates this to configuration files with the description of the experiments, ready to be used by the experiment controller. During the generation of the experiments, hints are generated for each workflow job, so that they can be matched directly with the anomalous resources if the model confidence of that job for the specified anomaly falls into the lower confidence regions. The translations layer is responsible for invoking the experiment controller and waiting until all experiments have finished. After the experiments' completion, the controller parses the logs and generates summaries of the statistics for each job in a CSV file - tabular data format, that is then passed back to the active learning module.

With the data-generating mechanism providing tasks corresponding to each iteration, we describe the complete algorithm next.
\subsection{Algorithm}
Note that the overall active learning algorithm is described in Algorithm~\ref{alg1a} and an overview is provided in Figure \ref{fig:overview}. In particular, we propose a search and minimize strategy. This strategy evolves in two steps. First,  we search for the best samples to accurately calculate the cost.  To perform this search, our algorithm interfaces with the experimental infrastructure provided by Poseidon-X, represented by the function $g$ in Algorithm~\ref{alg1a}. In particular, we generate the score $s_{k-1}$ and feed it to $g$ to obtain the sample $\cT_{k}$.  In the second step we minimize the cost $H$ corresponding to this sample using stochastic gradient descent based approach. These two steps are repetitively performed until convergence. 

Since the first step in this algorithm is a search strategy, the model can get unusually biased due to initial batches of data. Therefore, the use of burn-in is very important and a burn in is performed at the initial steps of the algorithm. Unfortunately, in any active learning type approach, this cost of burn-in must be paid and without it, the model starts off with a large jump in performance which slowly reduces because of increasing bias with samples. This phenomenon and the results of our approach are described in the next section.

\begin{algorithm}
    Initialize ${\vect{w}}, s_{0}$\;
    Initialize $\cT_{1} = g(s_{0})$\;
    Set BurnIn\;
    \For{$k=1,2,3,\ldots,K$}{
        \If{$k > BurnIn$}{
            $\cT_{k} = g(s_{0})$\;
        }
        \Else{
            $\cT_{k} = g(s_{k-1})$\;
        }
        Evaluate $H(\vect{w}, \cT_{k})$\;
        Minimization step: Update the weights $\vect{w}$ through stochastic gradient descent on $H(\vect{w}, \cT_{k})$\;
        \If{$k > BurnIn$}{
            Evaluate $s_{k-1}$\;
        }
        \If{$\|s_{k-1}\| \leq threshold$}{
            STOP\;
        }
    }
    \caption{Active Learning \label{alg1a}}
\end{algorithm}

\section{Experimental Results}~\label{sec:experiments}
In this section, we will demonstrate that the active learning algorithm described in this paper can effectively minimize uncertainty for any given model and improve the learning behavior of that model. To demonstrate this improvement we will choose a total of three workflows, 1000Genome, Montage, and Predict Future Sales. Moreover, we will show the efficiency of two different model classes-- the supervised learning case, where we assume that we have labels for all the data; and the self-supervised learning model, where assume the lack of labels.

Using these workflows and the model, we utilize two experimental settings. First, for the 1000Genome workflow, we will demonstrate an end-to-end active learning experiment where the feedback from our approach is utilized for running experiments using Poseidon-X (Section~\ref{sec:experimental-infra}). Second, for the Montage and the Predict Future Sales workflow, we will describe an emulation for the active learning framework, using data offered by the Flow-Bench benchmark dataset~\cite{papadimitriou2023flowbench}, and show that the experimental advantages are carried over to the emulation as well. We will begin by describing the metrics and the baseline we use in the evaluation. 

\paragraph{Metrics}
In this paper, we measure performance through three metrics: ROC-AUC score, average precision score, and top-k precision. In particular, the ROC-AUC score describes the models' ability to distinguish between different classes by evaluating ``how many data-points have been correctly classified into their respective classes?" On the other hand, the average precision score indicates the ability of the model to correctly identify all the positive examples without accidentally identifying false positives. Top-k precision quantifies the proportion of the correct label among the top k labels predicted by the model~\cite{petersen2022differentiable}. For the sake of calculating the metrics, we hold out a test data set from previously collected data regarding all the different workflows considered in this paper. All our experiments, including the end-to-end active-learning experiment, are repeated for five independent random initializations.

\paragraph{Baseline:}
For each of the three workflows, we will develop two baseline models, one using the self-supervised learning~(SSL) approach described in~\cite{papadimitriou2023flowbench} and the other using the supervised learning model described in \cite{jin2023graph}. 

\subsection{Self-supervised Learning Module} \label{ref:ssl}
\begin{figure*}
    \centering
    \includegraphics[width=0.9\textwidth]{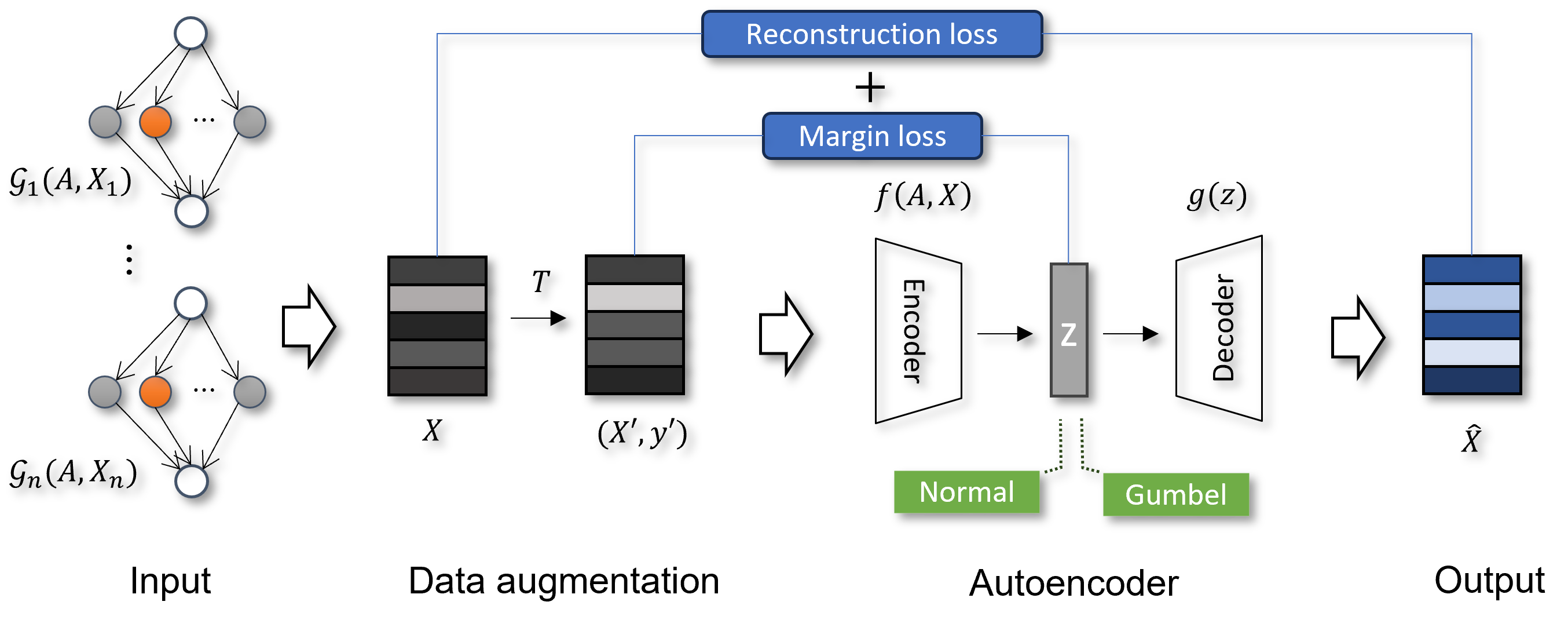}
    \caption{A pictorial representation of the self-supervised learning model~\cite{jin2023ssl}. The SSL model is comprised of a graph driven encoder and a multi-layer perceptron-based decoder.}
    \label{fig:SSL_diagram}
\end{figure*}

The self-supervised learning approach is based on the unsupervised learning paradigm and is very important because, there is no guidance of labels to train the model. To circumvent the lack of labels, self-supervised learning (SSL) creates target objectives without human annotation. SSL models have a two-step process. A pictorial representation is given in \ref{fig:SSL_diagram}.

First, data augmentation and transformation generate a modified version of the input graph. In particular, the workflow data, modeled as a directed acyclic graph is provided as the ``input graph" and augmentations procedures and transformation detailed in \cite{jin2023ssl} are applied to attain a modified version of the ``input graph." Since each input graph is comprised of a feature matrix and an adjacency matrix, for each input graph an augmented feature matrix is obtained. Moreover, for this modified input graph, a set of augmented labels is defined. This augmented feature matrix paired with the adjacency matrix and the augmented label describes the ``modified version of the input graph." This pair of input graphs (the original and the modified) is fed into the SSL model. 

In the SSL model, an encoder is applied to the two graphs and these graphs are converted into the latent space. In the latent space, the distribution of the normal and anomaly jobs is captured by the Gumbel distribution~\cite{jin2023ssl} with the Gumbel \texttt{softmax} approach for sampling. The samples from the latent space are then fed into a decoder which is constructed using a multilayer perceptron. 

To train this model, a combination of reconstruction error and margin loss is utilized, precise details are provided in \cite{jin2023ssl}. 

\subsection{Supervised Learning Module} \label{ref:sl}
\begin{figure*}
    \includegraphics[width=\textwidth]{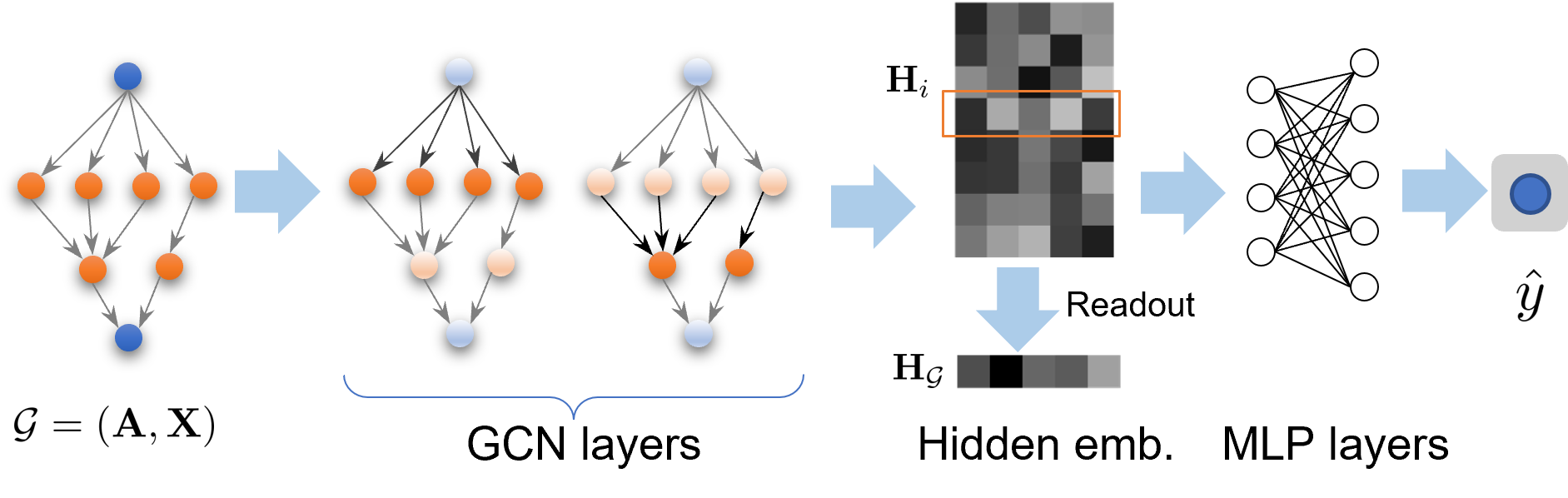}
    \caption{This is a pictorial representation of the 
 supervised learning-driven graph neural network module~\cite{jin2022workflow}. This module is comprised of 2 graph convolutional layers appended with a multi-layer perceptron.}
\end{figure*}
The supervised learning model is composed of a standard graph convolutional network module with multi-layer perceptron at the end . We utilize the cross entropy loss function for the supervised learning problem and the stochastic gradient descent for learning. Additional details regarding the network can be found on \cite{jin2022workflow}.

For both these models, we use the AdamW optimizer~\cite{reddi2019convergence}. For all three application workflows, we use $5\times10^{-4}$ for the learning rate, and $10^{-4}$ for weight decay, and a batch size of 32. We use the models identical to the ones described in \cite{papadimitriou2023flowbench}  and \cite{jin2023graph}. For the baseline run, we extract 28 samples for the 1000Genome workflow, 100 for Montage and 250 samples for Predict Future Sales workflow. Each sample is comprised of one batch of data. Each data-point in this batch of data is essentially a directed acyclic graph, comprised of nodes~(jobs) and edges, therefore, we will refer to this batch of data as ``workflow graphs". The model (either the SSL one or the supervised learning on) is updated exactly once using the workflow graphs. In the following results, we wish to determine, ``how fast the model converges and what are the minimum number of workflow graphs required to train a model to acceptable accuracy?"

\subsection{End-to-end Active Learning Experiment}
In this test case, we integrate our active learning framework with the Poseidon-X experimental infrastructure (Section~\ref{sec:experimental-infra}) where we run experiments guided by our algorithm. In this online setup, the workflow graphs is obtained by introducing anomalies on the fly. In particular, our uncertainty score feeds requirements on needed workflow graphs to the controller and the controller executes the workflows and feeds the data back into the model for further training. We execute this two-step strategy for 28 runs such that 28 different batches of data are extracted. We run two end-to-end active learning experiments with this construct -- one with the supervised learning model and one with the self-supervised learning model. The experimental settings for this are provided as follows.

\paragraph{Experimental setting~(End-to-end Interface)}
  We execute a total of 10 iterations of data generation, with $2$ workflow executions per iteration. These executions are performed by interfacing the scores generated by our active learning methods with the experimental controller described in \ref{sec:experimental-infra}.

  We utilize the following settings for the active learning part of the experiments.
 \paragraph{Active-learning settings}  To execute our self-supervised learning model, we divide the total number of workflow graphs~(the batch of data provided by the Poseidon-X infrastructure) into batches of $32$ nodes~(a collection of 32 jobs/compute tasks) each. We pre-process the workflow graphs generated by the experiment to be suitable for a node classification problem. Specifically we accumulate the workflow graphs into a large graph utilizing a block diagonal adjacency matrix. The process of this pre-processing can be found in~\cite{jin2023ssl}. For each set of workflow graphs, we update the model exactly once and record the ROC-AUC, precision, and $top-k$ score for each of these executions. Note that, for all the comparisons, we keep the model definitions and other parameters identical between the baseline and our active learning experiment. With identical model architectures, we observe the change in the test metrics with and without the active learning component.  
\begin{figure*}
    \centering
    \includegraphics[width = 0.7\linewidth]{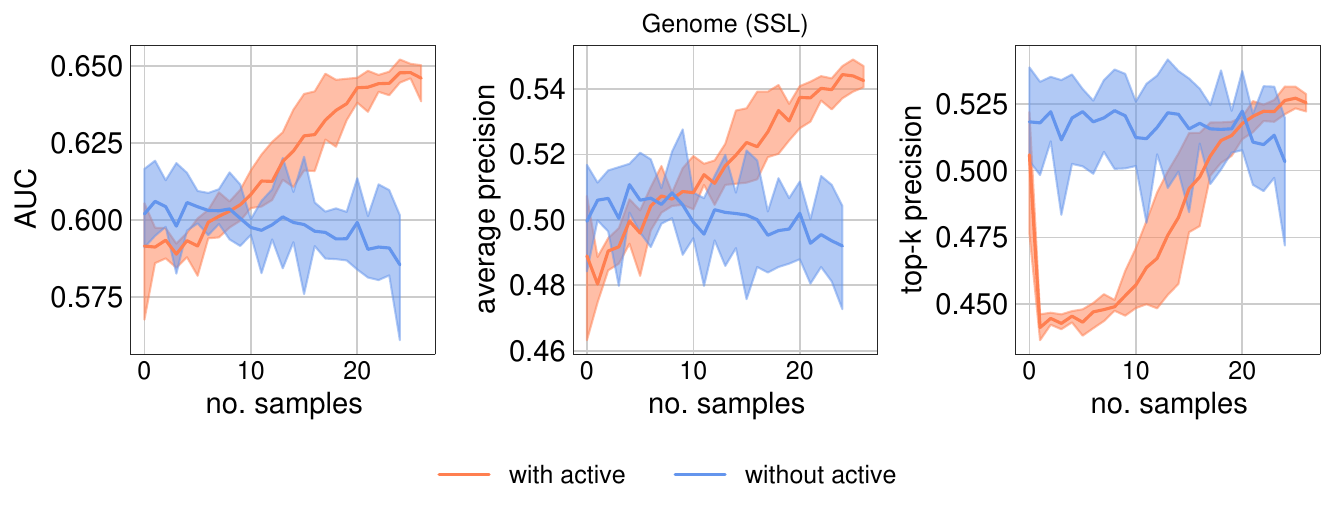}
    \includegraphics[width = 0.7\linewidth]{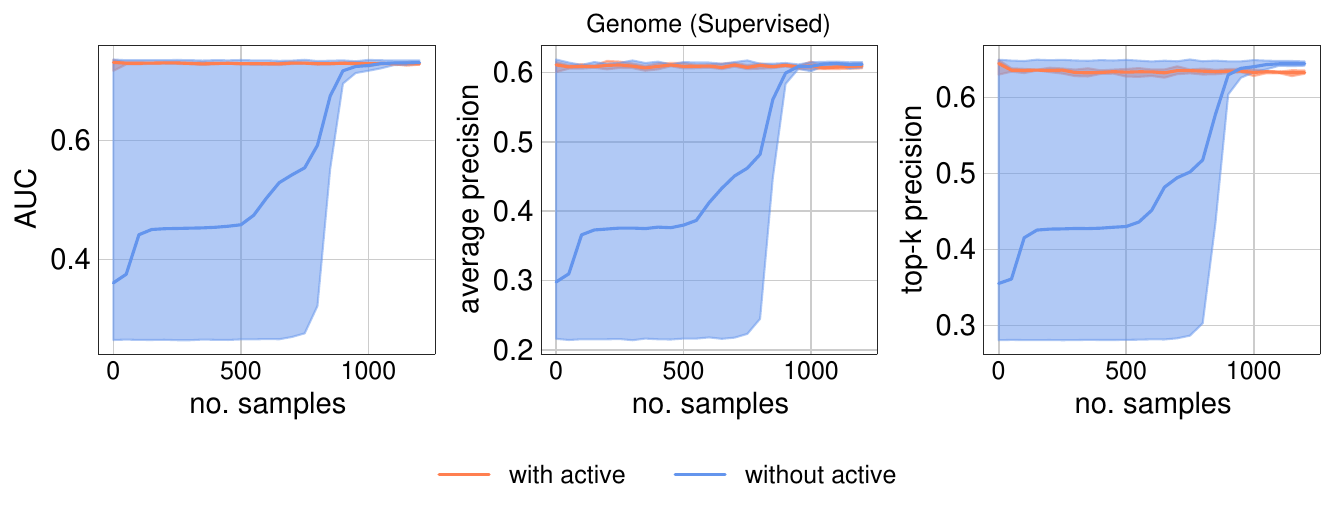}
    \caption{ Training on 1000Genome workflow with a full-scale end-to-end experiment with and without active training. }
    \label{fig:experiment__results}
\end{figure*}

\paragraph{Results:} We present the results of this experiment in Figure~\ref{fig:experiment__results} where we observe the progression of the AUC, precision, and top-k as a function of a number of samples. At the top, we show the results with the self-supervised model, and at the bottom, we show the results with the supervised learning model.  Since, at each iteration, there is a total of two samples, the total number of samples generated is twenty. For each sample in the case of 1000Genome workflow, a total of $138$ jobs are executed by the experimental controller. Therefore, for each learning iteration, Poseidon-X provides data for $2*138$ jobs. First, we note that for the active learning case, the first iteration involves randomly generating data because, at the first iteration, there is no information regarding the uncertainty of the model, due to this, both the active and the non-active cases exhibit similar values of AUC, precision and top-k scores for the different metrics. However, immediately after the first iteration, the active learning method goes through an exploration phase and this results in a small dip in performance. However, as the learning progresses, the active learning approach progressively improves its uncertainty and performs better. In fact, while the standard learning plateaus due to a lack of data (only 56 samples are provided to the standard learning model), the active learning approach significantly outperforms the standard one. 

This outperforming behavior is even more apparent in the supervised learning scenario. First of all, it is expected that supervised learning models achieve higher AUC and precision compared to the SSL model, this is evident from the bottom plots in \ref{fig:experiment__results}. Next, we see that the active learning approach is significantly outperforming the standard supervised learning case. Moreover, there is significant uncertainty in the performance as observed by the large blue bands observed in the standard supervised learning scenario. This uncertainty is significantly reduced by the active learning approach, where the orange curve has a very small band.  One particularly important thing to note here is that these results do not imply that the standard learning scenario cannot perform well on the 1000Genome workflow. However,  for the given amount of data samples, the standard learning scenario is significantly outperformed by the active learning approach where data is requested based on the uncertainty of the model's performance on the data.  

\subsection{Active learning Emulation}
Due to resource constraints, we perform an active learning emulation for the other two workflows, i.e.-- the Montage and the Predict Future Sales workflow.  In this case, we reuse the workflow performance data obtained via prior experiments.

\paragraph{Experimental Setting:} At each iteration of the end-to-end active learning experiment, we generate a total of 2 workflow graphs for the Montage and Predict Future Sales and the learning performed on these two graphs ($2\times 138$) per iteration. These graphs are generated by determining the what type of anomalies should be injected to which jobs which is determined by selecting anomalies corresponding to the low confidence region of the model. To emulate this behavior and keep fidelity with the active learning, we develop a strategy where we select the anomaly with the lowest confidence value and use that to generate the next batch of data. It must be noted that, for the live experiment, multiple jobs in a graph can be introduced with distinct anomalies. However, in the emulation scenario, this is not feasible as the emulation strategy selects jobs corresponding to just one anomaly at a time. This has the potential to bias the model unduly towards this anomaly. Therefore, for every few iterations, we select the anomaly at random to reduce this bias in the experiment. The rest of the settings from the active learning end-to-end experiment are transferred to the emulation setup.

\begin{figure*}
    \centering
    \includegraphics[width = 0.7\linewidth]{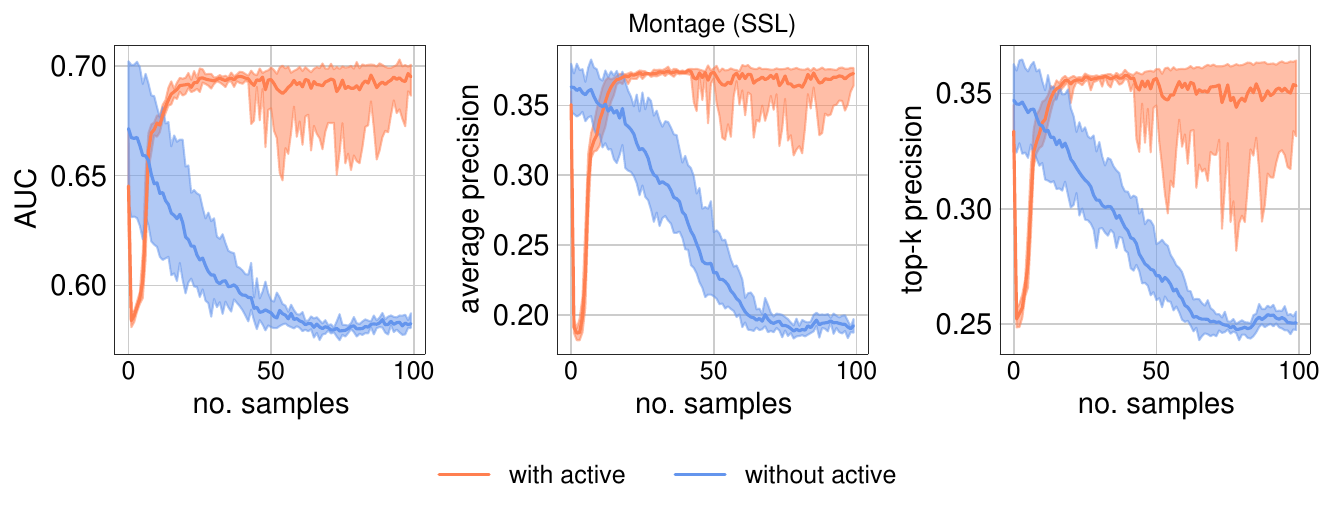}
    \includegraphics[width = 0.7\linewidth]{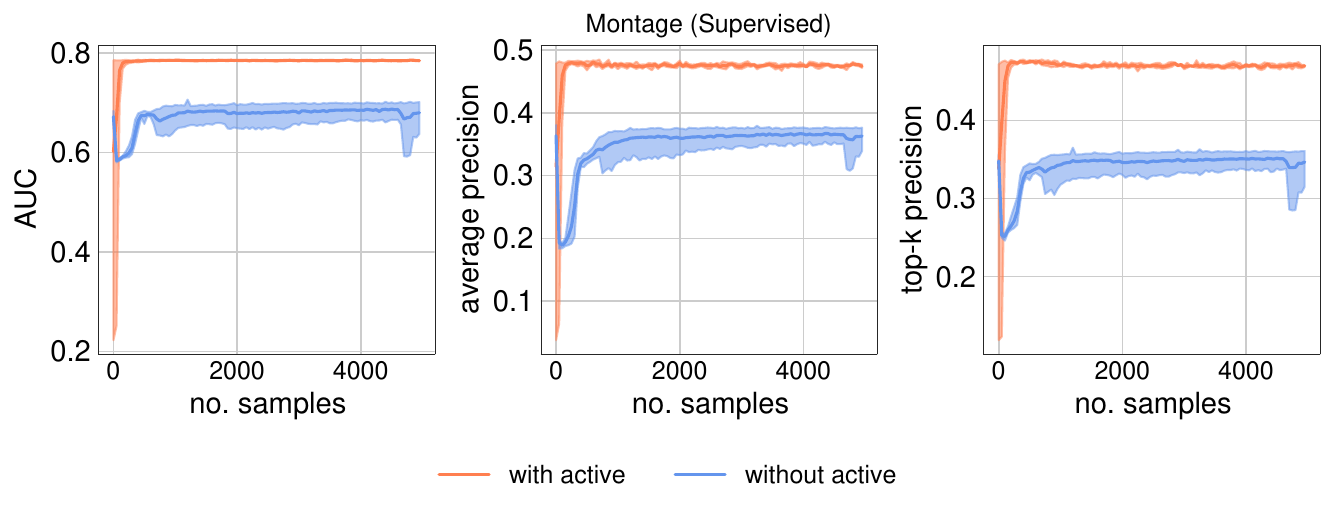}
    \caption{Training on Montage workflow with and without active learning.}
    \label{fig:montage}
\end{figure*}

\begin{figure*}
    \centering
    \includegraphics[width = 0.7\linewidth]{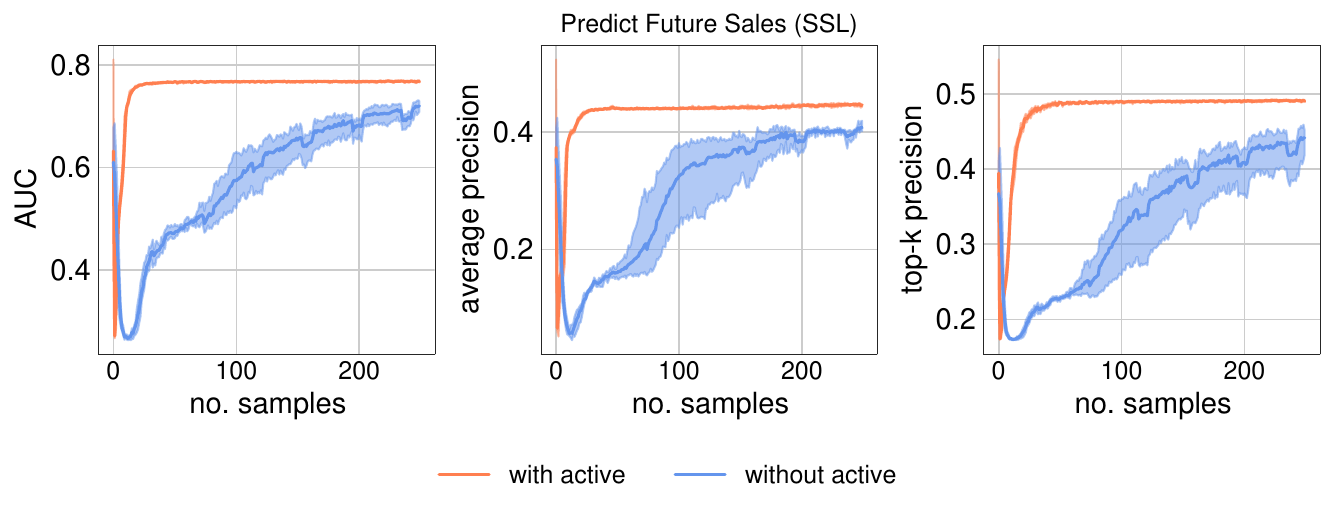}
    \includegraphics[width = 0.7\linewidth]{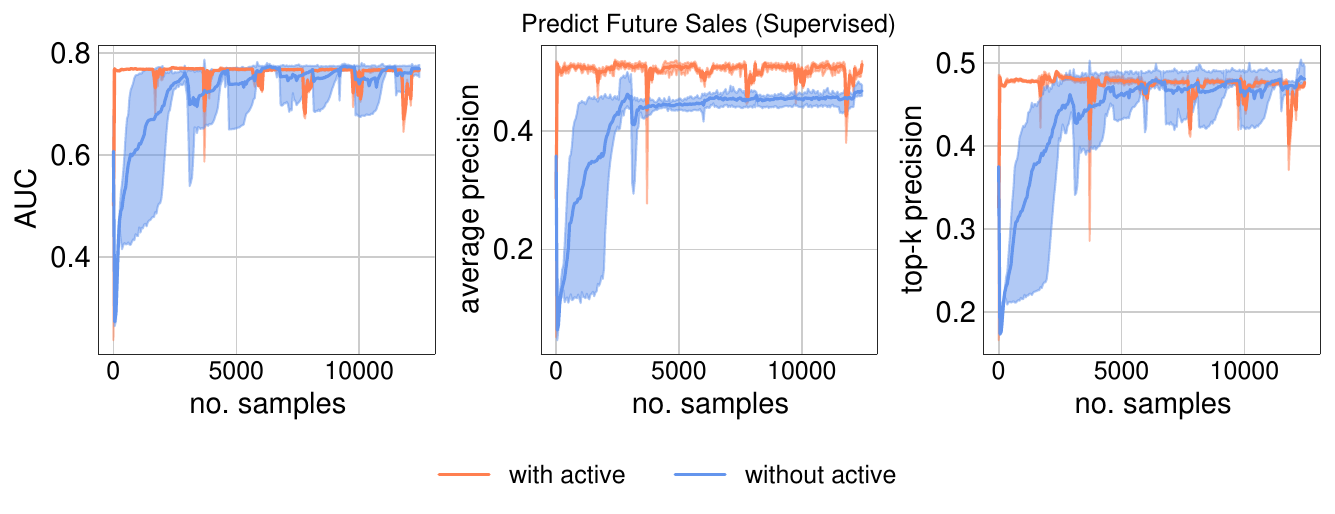}
    \caption{Training on Predict Future Sales workflow with and without active learning.}
    \label{fig:predict_future_sales}
\end{figure*}

\paragraph{Results:}
We observe that the trends from the end-to-end experiment, that uses the 1000Genome workflow (Figure~\ref{fig:experiment__results}), are being repeated in the active learning emulation experiments that use precaptured data from the Montage workflow (Figure~\ref{fig:montage}) and the Predict Future Sales workflow (Figure~\ref{fig:predict_future_sales}).
For the Montage workflow (Figure~\ref{fig:experiment__results}), we observe that the standard model and the active learning model start at similar levels of AUC, precision, and top-k. Next, there is a dip in the performance during active learning as expected in the exploration phase of the learning. Then, the performance of the actively learned model improves much faster compared to the non-active learning case. This behavior is also observed in the Predict Future Sales workflow (Figure~\ref{fig:predict_future_sales}). Additionally, for the Montage workflow, we see a steady decrease in the performance of the SSL model. This is a common learning behavior for both the SSL and the supervised learning models and with increased training (more epochs and data), the model reaches back to the precision levels obtained by the active learning approach. We see this behavior in the bottom plot of Figure \ref{fig:montage}. As mentioned earlier, the main advantage of the active learning paradigm is that higher accuracies are achieved with a significantly fewer number of samples relative to the standard learning case as is the case in the supervised learning scenario in Figure \ref{fig:predict_future_sales}. For the Predict Future Sales workflow, we observe that the results for the active learning emulation are higher compared to the baseline. This improvement is about 15 \%. 

There are two main causes of performance improvement with our active learning paradigm. The first is due to the use of an extra term (Equation~\ref{eq:approximation}) in the cost function that seeks to reduce the bias of the model to workflow graphs. Second, we use uncertainty values to drive anomaly injection in the workflow and improve the convergence aspects of the model. In what follows, we will describe ``why our approach behaves this way."

\subsection{Insights into the active learning}
\begin{figure*}
    \centering
    \includegraphics[width = 0.7\linewidth]{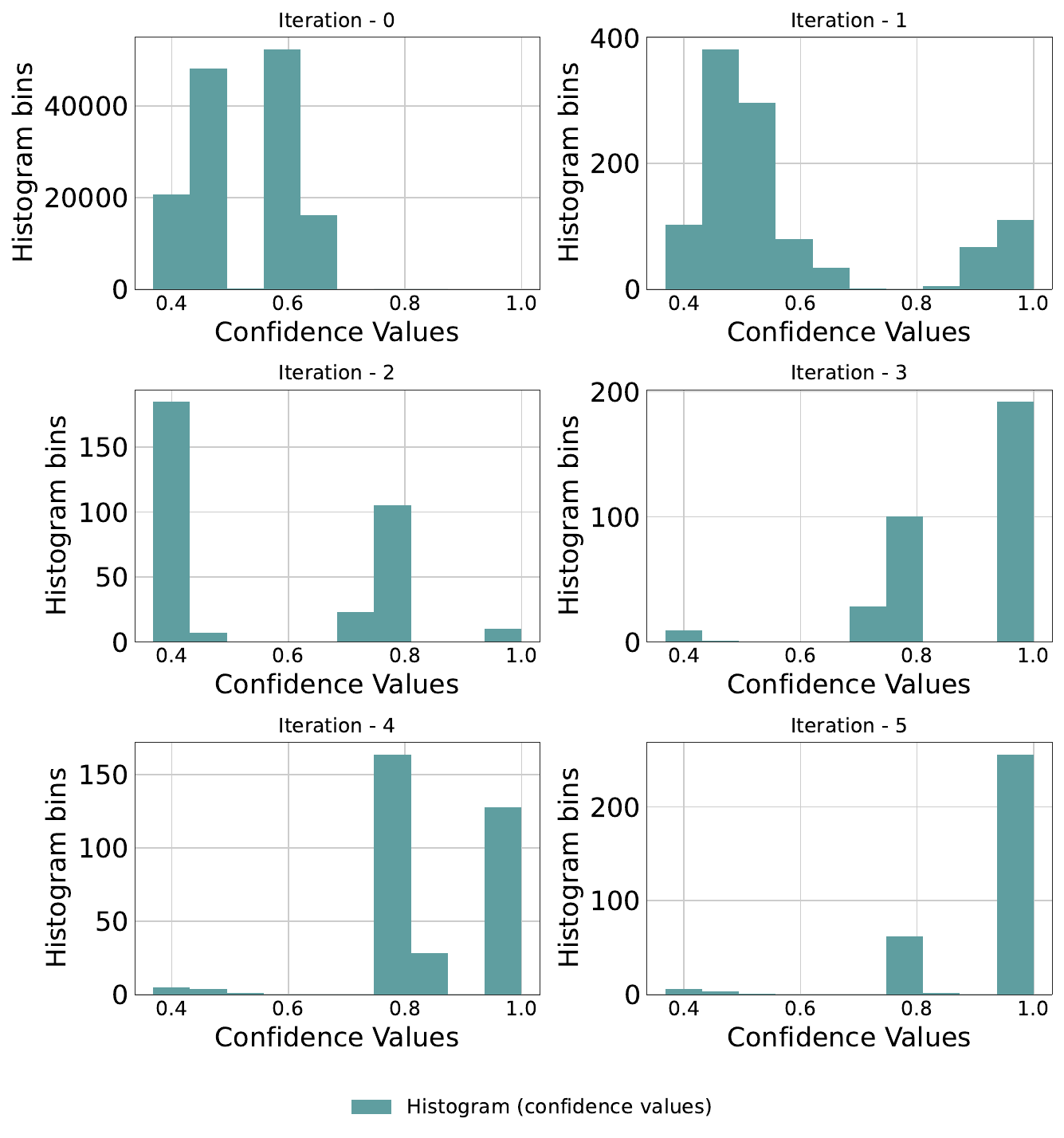}
    \caption{Histogram of the confidence values with respect to iterations.}
    \label{fig:hist}
\end{figure*}

It is important to note that convergence in the active learning aspect is related to the uncertainty scores. Towards this end, we plot the histograms of the uncertainty scores estimated at each iteration in Figure~\ref{fig:hist}. Note now that, the uncertainty scores are normalized between zero and one. Therefore, a score closer to one indicates that the model has high confidence in the data and the model output. On the other hand, a score closer to zero indicates that the model has low confidence. The goal of active learning is to nudge the histogram of uncertainty score towards one. That is, we expect the histograms of the uncertainty scores to be broad at the beginning of the learning process and very narrow at the end. This behavior is observed in Figure.~\ref{fig:hist}. At the start of the learning process, we see a very broad histogram of uncertainty where the values are closer to $0.5.$ Moreover, as the learning progresses, the histogram becomes narrower and closer to one. This indicates that the process of active learning is specifically reducing uncertainty and this results in better metrics. In particular, this minimization of uncertainty is happening because our loss function setup in Equation \ref{eq:approximation} has an extra term (3rd term) corresponding to the uncertainty of the model which facilitates minimizing uncertainty in the learning problem. 

\begin{figure*}
    \centering
    \includegraphics[width = 0.7\linewidth]{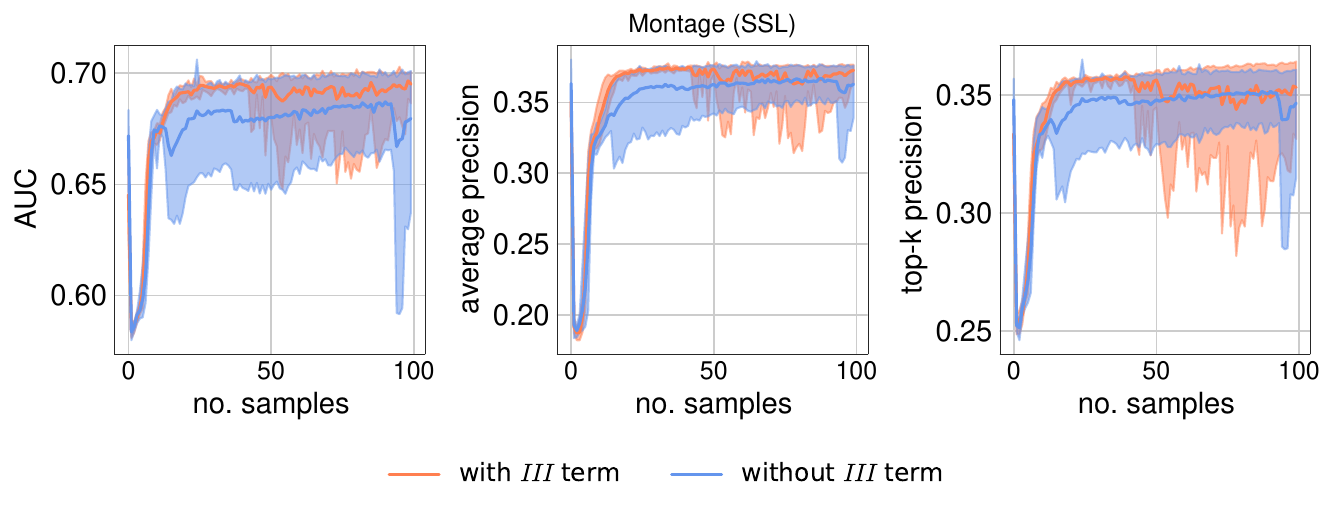} \\
    \includegraphics[width = 0.7\linewidth]{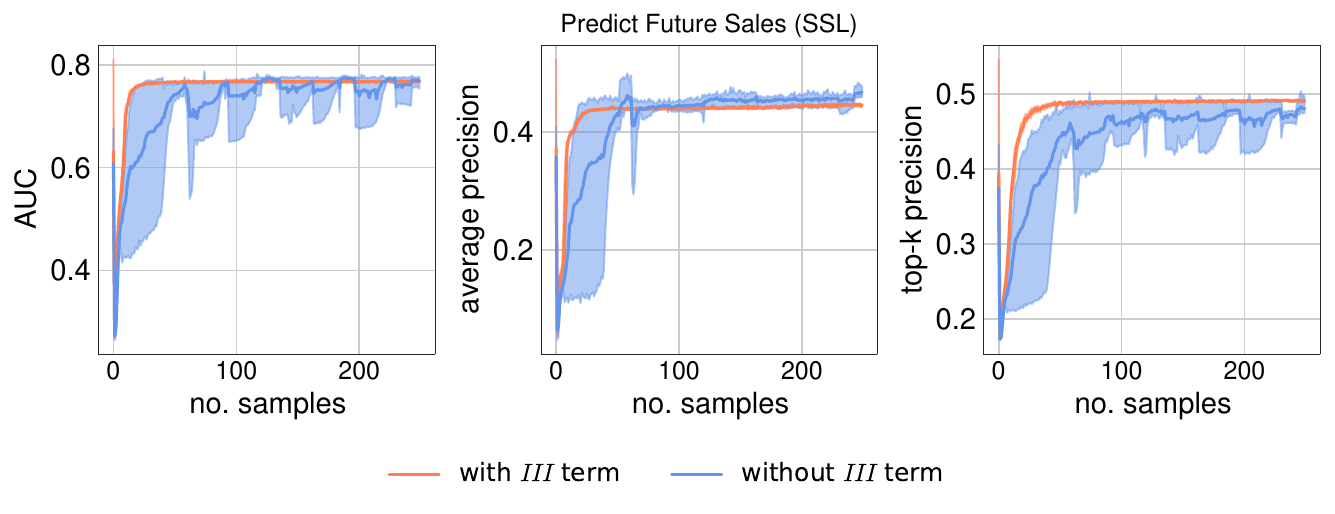}
    \caption{Training on the SSL models with and without the third term in Equation \ref{eq:approximation}.}
    \label{fig:SSL_withwithout3}
\end{figure*}

\begin{figure*}
    \centering
    \includegraphics[width = 0.7\linewidth]{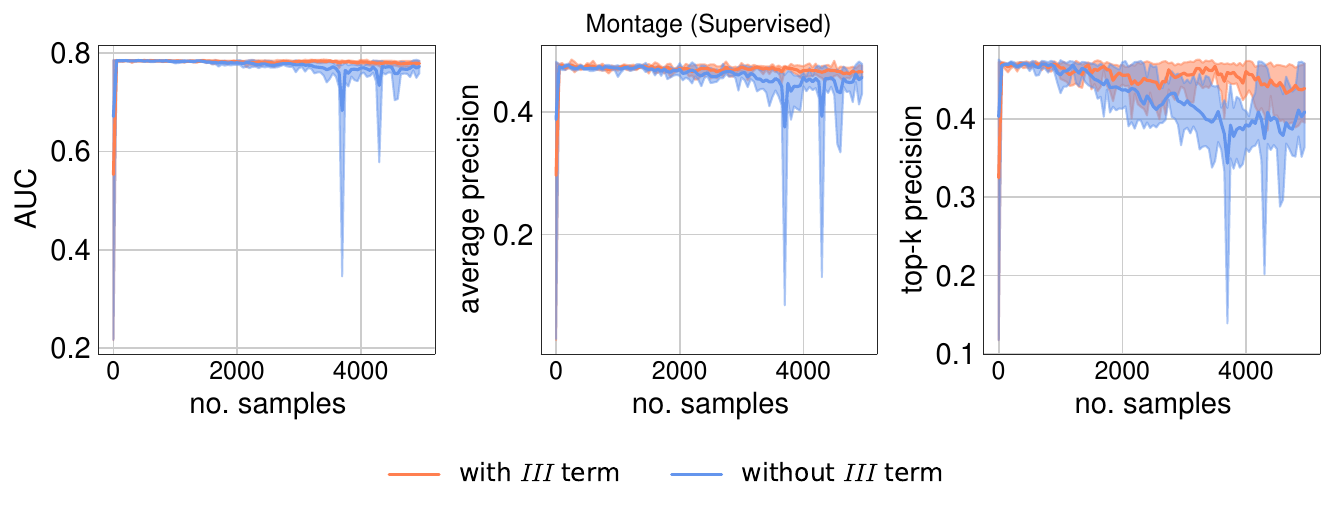} \\
    \includegraphics[width = 0.7\linewidth]{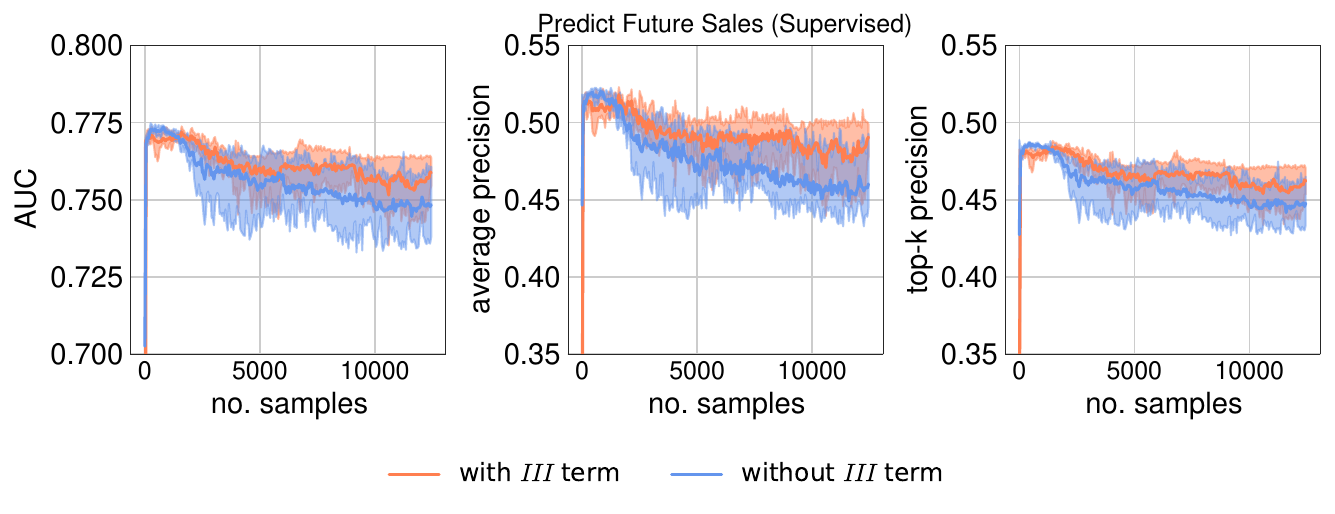}
    \caption{Training on the supervised learning model with and without the third term in Equation \ref{eq:approximation}.}
    \label{fig:supervised_withwithout3}
\end{figure*}

Next, we will analyze the effect of this extra term in the loss function in Equation \ref{eq:approximation} by plotting the learning process with and without the uncertainty term in the active learning setup. The progression of this learning with respect to the number of samples is shown in Figure.~\ref{fig:SSL_withwithout3} and \ref{fig:supervised_withwithout3}. In these figures, we observe that without the additional term, the evaluated metrics are much worse than the case when the additional term is present for both the SSL model and the supervised learning model. Furthermore, in supervised learning case, the number of iterations involved in the updates is much larger compared to the SSL scenario. In this case, a deteriorating behavior is observed, wherein the curves without the uncertainty term (the third term in Equation \ref{eq:approximation}) deteriorate more relative to the case with the third term. In ML literature, this behavior is a sign of over-training. In this case, what is happening is that the model is increasing its bias with each successive iteration as seen in the case of the Predict Future Sales workflow in Figure.~\ref{fig:supervised_withwithout3}, where the score decreases with increased training. However, the deterioration is much smaller compared to the case with the additional term in Equation \ref{eq:approximation} and this behavior can be tuned through the optimization.

The aforementioned results cement the conclusion that explicitly leveraging uncertainty to improve sample generation provides better learning and reduces the need for significant resources.

\section{Conclusion}
\label{sec:conclusion}
In this work, we presented an experimental framework, Poseidon-X, that leverages the Pegasus workflow management system and two NSF-funded cloud testbeds (Chameleon and FABRIC), to provide reproducible and on-demand data collection of computational workflow executions. Poseidon-X supports data collection from any workflow, after it is ported into a Pegasus workflow. Additionally, by using a custom isolated deployment across the two cloud testbeds, Poseidon-X supports injection of different types of anomalies (CPU, HDD, Network) at variable magnitudes, which are tracked automatically and are correlated with the workflow executions to generate labeled datasets.
Leveraging this on-demand experimental framework we developed an active learning approach for computational workflows, that allows the Machine Learning models to drive the creation of new samples and gather new training data points in order to improve their predictions.
We demonstrated that the active learning approach can accelerate the ML model training for anomaly detection on computational workflows and we showcased this using an end-to-end live experiment and through emulation using real pre-captured data.
Our experiments highlight that generating data based on the needs of the model under training, improves the training behavior and reduces the amount of resources consumed to create and harvest new data points from computational workflows.
In the future we are planning to expand this framework to be adaptable to any type of workflow and explore how transfer learning can further improve the active learning methodology.

\section*{Acknowledgment}
This work is supported by the Department of Energy under the Integrated Computational and Data Infrastructure (ICDI) for Scientific Discovery, grant DE-SC0022328. Additionally, this work is supported by the U.S. Department of Energy, Office of Science, under contract number DE-AC02-06CH11357. Finally, all development and experimentation was conducted on Chameleon Cloud and FABRIC Testbed, and we want to thank the testbeds for their support.

\bibliographystyle{elsarticle-num}
\bibliography{references}

\end{document}